\documentclass[twocolumn]{aa}
\usepackage{graphicx}
\usepackage{txfonts}
\usepackage{natbib}
\begin{document}
   \title{2MASS NIR photometry for 693 candidate globular clusters in M31 \\
and the Revised Bologna Catalogue \thanks{Tables 2, 3 and 4 are only
available in electronic form at the CDS via anonymous ftp to 
{\tt cdsarc.u-strasbg.fr} (130.79.128.5) or via 
{\tt http://cdsweb.u-strasbg.fr/cgi-bin/qcat?J/A+A/n/n}}
}

   \author{S. Galleti\inst{1,2},
          L. Federici\inst{1},
          M. Bellazzini\inst{1},
          F. Fusi Pecci\inst{1},
          S. Macrina\inst{2}}

   \offprints{S. Galleti}

   \institute{INAF - Osservatorio Astronomico di Bologna, Via Ranzani 1,
        40127 Bologna, Italy
               \and
             Dipartimento di Astronomia, Universit\`a di Bologna,
              Via Ranzani 1, 40127 Bologna, Italy\\
              \email{galleti,luciana,bellazzini,flavio@bo.astro.it}}
   \authorrunning{S.Galleti et al.}
   \titlerunning{2MASS NIR photometry for 693 globular clusters in M~31}
   \date{Received November 5, 2003; accepted December 2, 2003}

   \abstract{ We have identified in the 2MASS database 693 known and
candidate globular clusters in M~31. The 2MASS $J,~H,~K$ magnitudes of
these objects have been transformed to the same homogeneous
photometric system of existing near infrared photometry of M~31
globulars, finally yielding $J,~H,~K$ integrated photometry for 279
confirmed M~31 clusters, 406 unconfirmed candidates and 8 objects with
controversial classification. Of these objects 529 lacked any previous
estimate of their near infrared magnitudes.  The newly assembled near
infrared dataset has been implemented into a revised version of the
Bologna Catalogue of M~31 globulars, with updated optical ($UBVRI$)
photometry taken, when possible, from the most recent sources of CCD
photometry available in the literature and transformed to a common
photometric system. The final Revised Bologna Catalogue (available in
electronic form) is the most comprehensive list presently available of
confirmed and candidate M~31 globular clusters, with a total of 1164
entries.  In particular, it includes 337 confirmed GCs, 688 GC
candidates, 10 objects with controversial classification, 70 confirmed
galaxies, 55 confirmed stars, and 4 H~{\sc ii} regions lying
within $\sim 3\degr$ from the center of the M~31 galaxy. Using the
newly assembled database we show that the $V-K$ color provides a
powerful tool to discriminate between M~31 clusters and background
galaxies, and we identify a sample of 83 globular cluster candidates,
which is not likely to be contaminated by misclassified galaxies.

}

   \maketitle
%

\section{Introduction}

Globular clusters (GCs) are among the oldest astrophysical objects
known and they are found in galaxies of all morphological types. Thus,
globular clusters systems are extensively studied to recover
fundamental information on the formation and early evolution of their
host galaxies \citep{h98,Ashz}.

With present-day technology globular clusters can be fully resolved into stars
down to below the Main Sequence (MS) Turn Off (TO) only if they reside in the
neighbourhood of our own Galaxy \cite[i.e., within $\sim 150$ kpc,
see][]{fornax}. The evolved populations [Red Giant Branch (RGB) and Horizontal
Branch (HB)] may be resolved in the  outer regions of globulars out to $\sim 1$
Mpc with the Hubble Space Telescope \cite[see][and references therein]{rich},
and a handful of bright RGB stars have been resolved by \citet{harri} in a few
clusters of the galaxy NGC~5128,  at $\sim 4$ Mpc \cite[but see][for the
improvements achievable with the new  HST-ACS instrument]{brown}.  Therefore,
to investigate the globular cluster system of even relatively nearby galaxies
we have to rely on their integrated properties, e.g., magnitudes,
colors and spectral indices \citep{h98}.

In this framework, the study of the GC system of the Andromeda galaxy (M~31) is
of special importance, since it is the only case in which the integrated
properties of globular clusters, observed in conditions similar to those 
typical of farther galaxies, can be sensibly compared with their actual
resolved  stellar populations. Hence, apart from the obvious interest in the
early history of the largest galaxy of the Local Group, the M~31 system is the
ideal testbed to check our ability to obtain correct estimates of metallicity
and ages of extragalactic globulars from their integrated colors and spectra.

The study of the M~31 GC system has a long and venerable history,
starting with the first catalogues of candidates published by
\citet{hubble} and \citet{Seyfert}. Since then, a number of authors
have contributed to the detection, classification and analysis of M~31
globulars \cite[see, e.g.][and references
therein]{Vetesnik,Sargent,Crampton,Battistini0,Battistinia,
Battistinib,Mochejska,Barmbya,Barmbyb,Barmbyc,Barmbyd}.  In addition to the
lists of candidates, these studies provided a large amount of
data, mainly optical colors and spectral indices \cite[i.e., 435
entries in the catalogue by][]{Barmbya}. Near Infrared (NIR)
$J,~H,~K$ photometry is available only for $\sim 260$ candidate
clusters in the $J$ and $K$ passbands, and just for 127 in $H$ from the
observations by \citet{Frogel,Sitko,Bonolia,Bonolib,
Cohen,Barmbya,Barmbyb}.

The Two Micron All Sky Survey \cite[2MASS,][]{cutri}, providing
$J,~H,~K$ photometry of 99.998 \% of the sky down to $K\sim 15.5$
(completeness $> 99$ \% for $K\le 14.3$), offers for the first time
the opportunity of obtaining homogeneous NIR photometry for {\em all}
the candidate GCs of M~31 within the limiting magnitude of the survey.

In the present paper we exploit this opportunity, providing
$J,~H,~{\rm and} ~K$ 
integrated magnitudes for 279 confirmed
clusters \footnote{Confirmation of the classification is usually
obtained by spectroscopic measures or by resolving the object into
stars \cite[see,
e.g.,][]{federicia,federicib,huchra91,Barmbya,racinea,racineb}}, 406
unconfirmed candidates, and 8 objects for which the classification is
controversial\footnote{Controversial objects are those that have been
classified as ``confirmed globulars'' by some authors and as
``confirmed galaxies'' or stars by others.}.  Near-IR photometry is
less sensitive to interstellar extinction with respect to the
classical optical colors and provides useful complementary information
that may help for instance to disentangle the age-metallicity
degeneracy \cite[see][and references therein]{kp1,puzia}.

The plan of the paper is the following: in Sect.~2 we present the
master list of candidate M~31 GCs we have used as a reference
catalogue (Master Catalogue), and we describe how the database of
existing optical and NIR photometry has been assembled from the
various sources in the literature; in Sect.~3 the actual search for
counterparts of the Master Catalogue objects in the 2MASS database is
described and discussed; Sect.~4 illustrates the power of the
optical-NIR color index $V-K$ in the discrimination between {\em bona
fide} globular clusters and background galaxies, in the range
$V-K<3.0$; in Sect.~5 the Revised Bologna Catalogue of candidate
globular clusters in M~31 is presented and described. The main results
of the present analysis are summarized in Sect.~6.


\section{The Master Catalogue}

A reliable and complete catalogue of M~31 globulars based on modern
CCD photometry has yet to be achieved even if limited to a quite
bright threshold $V=18$, i.e., $M_V \sim -7$ at the M~31
distance. This situation arises because of: (a) the large field of
view covered by the M~31 GC system (more than $9\times 9\deg$ in the
sky, if extended out to $\sim 100$kpc from the M~31 center), (b) the
possible contamination by foreground stars, other objects within M~31
itself, and background galaxies \cite[see][and references
therein]{Battistinia, Battistinib}, and (c) the variable apparent size
of the candidates (ranging from 2 up to 10 arcsec) often with
a strong and highly variable photometric fore/background due to
M~31 itself.  Therefore, the available catalogues which we have assembled
from different sources are necessarily incomplete and, to some
extent, contaminated. Moreover, both the photometric and the
spectroscopic data are generally not homogeneous.

As a first step, we compiled a master list of the known candidates
that has been reported in the literature up to 2003, June.  The list
of the original Bologna catalogue \citep{Battistinia} has been
implemented with (i) the lists of cluster candidates detected near the
nucleus by \citet{Battistinib} and \citet{Auriere}, (ii) the
candidates found by \citet{Mochejska} and, finally, (iii) the list of
new candidates found by \citet{Barmbyc} searching M~31 HST/WFPC2
images in the HST archive.

The final database (hereafter the Master Catalogue, MC) includes 1164
entries: 337 confirmed GCs, 688 candidate GCs, 10 objects with
controversial classification, 70 confirmed background galaxies, 55
confirmed stars, and 4 confirmed H~{\sc II} regions.  These
objects are spread out over an area of more than $3\degr \times
3\degr$ around the center of M~31. The positions of all the entries
are taken from the respective sources.  We decided to include also
previous candidates whose stellar or galactic nature has been
(subsequently) observationally ascertained. Information about these
objects may be useful (a) to avoid their re-discovery as candidate
globulars and (b) to characterize the typical population of sources
that may be misclassified as M~31 globulars.

\subsection{Optical Photometry}

The available photometric measures have been secured by the various
authors using different devices and techniques: CCD and photoelectric
photometry, photographic plates and, for several candidates, visual
photometry obtained by eye. To assemble all this material as
coherently as possible we ranked the sources of M~31 GCs photometry
according to the adopted technique, choosing CCD photometry whenever
available. Lacking CCD photometry, photoelectric photometry was
preferred to photographic photometry and the latter was preferred to
eye estimates based on photographic plates.  We chose to include
eye estimates to provide, at least, a guide for future
observations of the candidates which lack any instrumental
photometry. When one cluster was present in two different sets of CCD
photometry the source with the lower photometric uncertainty and/or
providing the larger number of entries was preferred.

In order to obtain a Master Catalogue with photometric measures as
homogeneous as possible we have taken as a photometric reference the
dataset by \citet[][their tab. 3]{Barmbya}, that gives homogeneous
$UBVRI$ data for 260 confirmed GC and candidates.  All other
catalogues considered in the present compilation have been transformed to
this reference list by applying the offsets we derived from the objects
in common between the considered catalogue and the list of Barmby et al.
\cite[see][for details]{macrina}. The derived offsets are reported in 
Table~1 and are in good
agreement with what found by \citet[][see their tab. 4]{Barmbya}.
 \begin{table*}[t]
\begin{center}
\caption{ Photometric Offsets}  
{

\begin{tabular}{llcccccccc} 
\hline  \hline
\multicolumn{7}{c}{} \\
Data&Catalog  & $\Delta$U          &  $\Delta$B       	&  $\Delta$V	    & $\Delta$R	   & $\Delta$I  & \\
\hline
\multicolumn{7}{c}{} \\
CCD:&         &         	       &  	 	        &  	 	         &	   &	    & \\
    &B93      & ....    	       & -0.37	 	        & 0.00	 	         & -0.42   & 0.00   &\\
    &RHH92/94 & ....    	       & 0.00	 	        & 0.00	 	         & 0.00	   & ....   &	 \\
    &M98      & ....    	       & ....	 	        & 0.00	 	         & ....	   & 0.00   &	 \\
PE: &         &         	       &  	 	        &  	 	         &	   &	    &  \\
    &SL83     & 0.00    	       &0.00	 	        & 0.00	 	         & ....	   & ....   &	 \\ 
    &SL85+    & 0.00    	       & 0.00	 	        & 0.00	 	         & ....	   & ....   &	\\
PG: &         &         	       &  	 	        &  	 	         &	   &	    &  \\
    &B87      & 0.00 (if U$\leq$17.9)  & 0.00  (if B$\leq$18)   & 0.00	 (if V$\leq$16.5)&	   &	    &  \\
    &         &-0.27U+4.8 (if U$>$17.9)&-0.31B+5.64 (if B$>$18) &-0.13V+2.17 (if V$>$16.5)&-0.03$r_{c}$+0.16&$\ast$  &  \\
\hline  \multicolumn{7}{c}{} \\
 \hline
\end{tabular}
}
{ 
\begin{flushleft}
Reference:
RHH92 \citet{reed92};
RHH94 \citet{reed94};
B93   \citet{Battistinib};
M98   \citet{Mochejska};
SL83   \citet{Sharov83};
SL85+ \citet{sh85} and succeeding papers; 
B87   \citet{Battistinia};($\ast$) unpublished I photographic magnitudes
      (Federici \& Fusi Pecci, private communication). 
     The indicated offset have been applied to the
     photometries of the different sources to transform them into the
     photometric system of \citet{Barmbya}.
\end{flushleft}
}
\end{center}
\end{table*}
In case of duplicated CCD observations from \citet{Barmbya} and
\citet{Barmbyc} the average value has been adopted. The same approach
has been adopted for duplicate photoelectric measures from
\citet{Sharov} and \citet{Sharov83}. No transformation is feasible for
the $V$ eye magnitudes of the 132 D class objects from
\citet{Battistini0} since this data set has too few objects (15) in
common with other catalogues to derive a reliable
transformation. 26 objects taken from the list of \citet{Crampton}
have no photometry from other sources, hence they lack optical
photometry in the Johnson-Cousins system and the magnitudes provided
by Crampton cannot be consistently transformed to the photometric
reference adopted here.

Concerning the photometric errors, one has to refer to the original
sources.  Typical internal errors for the CCD and photoelectric
magnitudes are less than 0.05 mag (0.08 in $U$); however, the
photometry of candidates affected by a strong background in the inner
M~31 regions typically has larger errors. For the photographic
magnitudes the typical errors range from 0.05 up to 0.20 mag, as shown
by the comparisons with the CCD data in common.  For the 15 objects
with both eye-estimates and CCD or photoelectric photometry, comparison of
$V$ magnitudes  suggests that the typical uncertainty of
the eye-estimates is $\pm0.5$ mag \citep{macrina}.

\subsection {Near IR Photometry}

Though the observations were carried out with different setups
(telescopes, filters, etc.), the existing NIR photometric data sets
\citep{Frogel,Sitko,Bonolia,Bonolib,Cohen} are in agreement to within
the errors \cite[see][]{Barmbya,Barmbyb}.  These errors are fairly
large: $\sim 0.1$ mag in $J$, and $\sim 0.2$ mag in $H$ and $K$.
Hence the $J,~H,~K$ magnitudes available before 2MASS can be
considered as a virtually homogeneous set. In the compilation of the
Master Catalogue, as a first step, we have therefore adopted for each
object the average values of all the available measurements, obtaining
$J$ and $K$ magnitudes for $\sim 260$ objects and $H$ magnitudes for 127
objects.

\subsection{Astrometry and Identifications}

The coordinates of the objects included in the MC are taken from the
original sources. They were measured with slightly different reference
frames and are thus affected by different internal and systematic
uncertainties. For this reason, performing the cross-correlation of
the MC with the 2MASS database (described in Sect. 3), we had to check
several cases of objects whose identification was not straightforward
(see below).  This analysis led to a few corrections of the original
MC that are described in the present section.

Several direct checks carried out during the preparation of HST and
spectroscopic observations \cite[see,
i.e.][]{Fusipecci96,rich03,Barmbyd} have confirmed that the average
astrometric accuracy of the coordinates in the \citet{Battistinia,
Battistinib} catalogue is better than $1\arcsec$.

For all A, B and confirmed C class objects from \citet{Battistinia}
with no 2MASS counterpart within $1\arcsec$ (see below) the coordinates 
have been checked on the Digitized Sky Survey (DSS-2) or on HST
archive WFPC2/STIS frames, when available.  This last check was needed
for the objects near the bulge or in regions with a very high
background, which cannot be seen on the DSS-2 images.  The MC
coordinates have been verified also for all the objects that had $JHK$
photometry in the MC, but that were not found in the 2MASS database
(see below).  The identification of these objects has been performed
directly on the available images and their listed positions has been
accordingly revised. 

In a few cases, we have found a discrepancy between the coordinates
reported in the \citet{Battistinia} catalogue and the corresponding
finding charts.  In particular, B472 is the brightest object $\sim
18\arcsec$ SE of what reported in the \citet{Battistinia} finding
chart, B453 is $\sim 27\arcsec$ NE, B450 is $\sim 1\arcmin$ SE, and
B284 is $\sim 3\arcmin$ S.  Moreover, the coordinates of B288 and B142
in \citet{Battistinia} are wrong, the correct values are reported in
Table 2.

In \citet{Battistinia} there are also a few wrong cross-identifications with
previous catalogues.    In table VII, G162 corresponds to SH11 and G171 to
SH12; as noted by  \citet{Sharov83} G162 and G171 actually are SH12 and SH11,
respectively. Moreover, in table IV of \citet{racinea} the object 231D-DAO61 
is actually 230D-DAO61 and in table V(b) DAO95 is DAO96.

Finally, the visual check carried out on the HST WFPC2/STIS archival frames
allowed a better  classification of a few candidates.  B123
and B196 are clearly resolved into stars by HST,  hence they have been
classified as confirmed globular clusters; B261 is  recognized as a foreground
star while B329 is a background galaxy. B445 and NB76 appear as good GC
candidates, while B430 is probably a star, but the classification is not
completely unambiguous in these cases. 

   \begin{figure}
   \centering
   \includegraphics[width=9.3cm]{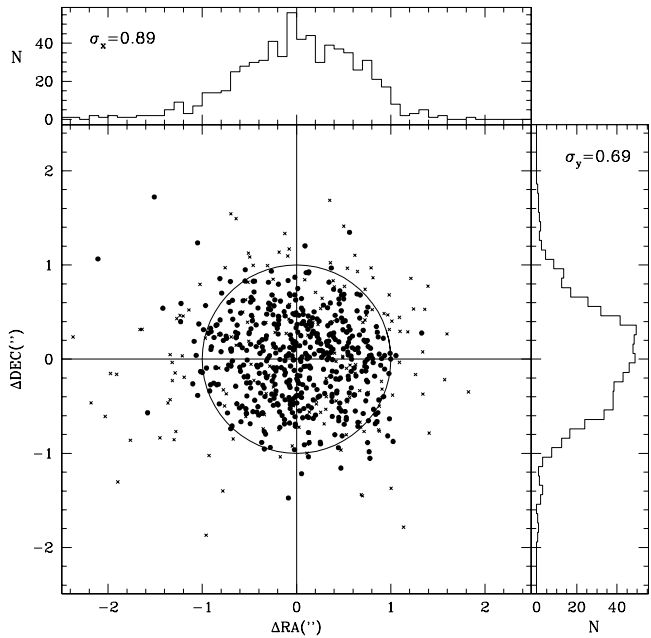}
    \caption{Residuals of the astrometric solution for PSC (filled circles)
    and XSC (crosses) sources. The histograms show the RMS of RA (top panel)
    and DEC (right panel). A circle of radius $r=1 \arcsec$ has been superposed
    to the plot as a reference. The standard deviation of the distributions
    of the residuals in each directions (histograms) are also reported within 
    the corresponding boxes.
    }
              \label{FigRes}%
    \end{figure}


\section{M~31 cluster candidates in the 2MASS database}

   \begin{figure}
   \centering
   \includegraphics[width=9.3cm]{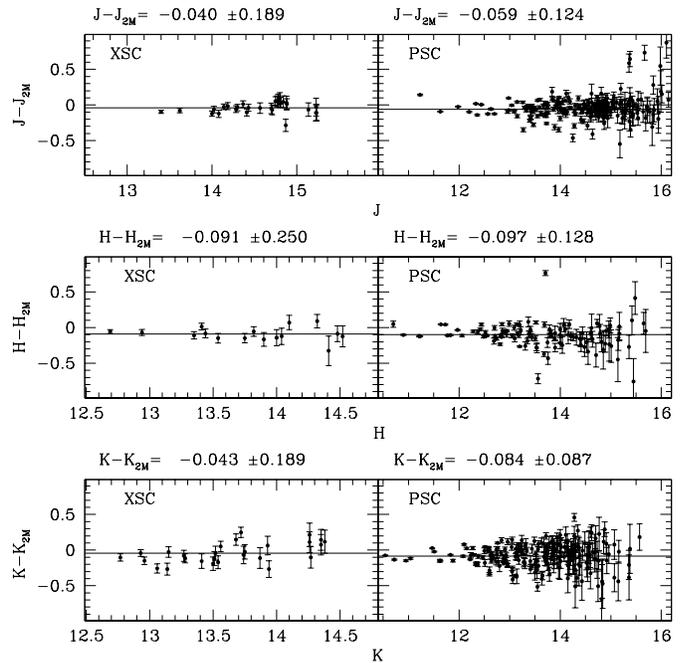}
    \caption{Comparison between the 2MASS photometry and previous
    NIR-photometry from the Master Catalogue. Left panels: 2MASS sources
    identifyied in the XSC. Right panels: 2MASS sources
    identifyied in the PSC.
    The extended source are in left panel and the points source in
    righ panel.
}
              \label{Fig2MS}%
    \end{figure}


   \begin{figure}
   \centering
   \includegraphics[width=9.3cm]{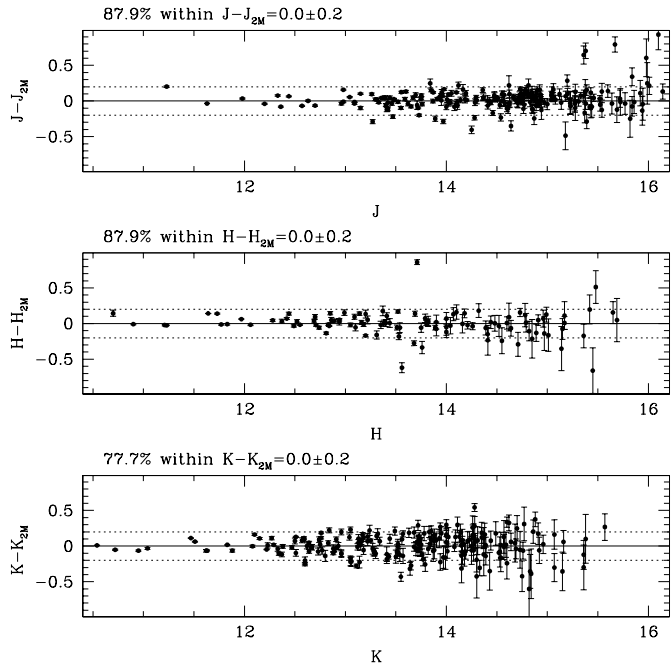}
     \caption{Comparison between the photometry of the final sample (corrected
     PSC+XSC) and previous NIR-photometry. The dashed lines encloses
     the range of $\pm 0.2$ mag.
               }
          \label{Fig2MT}
   \end{figure}
\subsection{2MASS IR photometry of the objects}

The classification of unresolved globular clusters is a very hard
task. Because of the round and compact appearance of unresolved GCs at
the typical resolution of ground-based images ($\sim 1 \arcsec$), both
distant galaxies and foreground stars can be misclassified as
GCs. Given the low spatial resolution of the survey, the automated
classification pipeline of 2MASS has assigned the M~31 GCs either to
the Point Source Catalogue (PSC) or to the Extended Source Catalog
(XSC), depending on their apparent sizes. Therefore we decided to
search the objects of the MC in both catalogues.

We selected from the 2MASS All-Sky Data Release all the sources within
an area of $4\degr \times 4\degr$ centered on M~31 from both the PSC
and the XSC data-sets, excluding all entries flagged as low-quality or
contaminated by nearby sources.  A general astrometric transformation
between the MC and the 2MASS catalogues was found using the many
entries in common as reference. The astrometric solutions were refined
in repeated iterations until a typical residual $< 1\arcsec$ was
achieved.  The final cross-correlation between the MC and the PSC
provided 597 identifications, while 179 MC objects have been
cross-identified in the XSC. In both cases the identification was
considered safe if a 2MASS source was found within $1 \arcsec$ from
the position of a MC entry (see Fig. 1). The identification of the sources
found at distances larger the $1 \arcsec$ from one MC entry has been checked by
visual inspection and has been accepted only if there was no possible ambiguity.
Note that most of the identifications with residuals larger than $1 \arcsec$
regards XSC entries.  
Of the total of 776 identified objects,
693 are confirmed or candidate globular clusters, the remaining 83 are
confirmed stars, H~{\sc ii} regions or galaxies\footnote{The reported number 
of MC objects actually identified in the 2MASS databases depends on 
the considered passband. For example the total number of confirmed and candidate
clusters who have a couterpart in 2MASS are 693 in J, 675 in H and 620 in K;
the number of objects who lacked any previous IR photometry are 529, 637 and 463
if we consider the J, H and K passband, respectively. For brevity and to avoid
confusion, all over the paper we report only the numbers referring to the 
J band measures. The corresponding numbers in the other passbands are similar,
the exact numbers can be obtained by inspecting the catalogues presented in 
Tab. 2, 3 and 4}.  More than one
hundred of the confirmed or candidate GCs identified in the 2MASS
databases have previous NIR photometry. In particular, 247 objects
have previous $J$ photometry, 122 have previous $H$ photometry and 236
have previous $K$ photometry. This large set of objects in common
allows a sensible comparison between the different photometries (see
below).  There are $\sim 400$ MC entries for which we cannot find a
counterpart in the 2MASS catalogues. These include $\sim 20$ (the
exact numbers depends on the considered passband) objects with
pre-2MASS NIR photometry.  Almost all of these are fainter than the
limiting magnitude of the 2MASS survey. Finally, the MC contains 88 entries
\cite[from different sources,
e.g.][]{Battistinia,Battistinib,Crampton,Auriere} that have no
optical photometry. These include 7 confirmed GC, 56 GC candidates and
25 confirmed stars/galaxies/H~{\sc ii} regions. We provide here $J,~H,~K$
photometry from 2MASS for 30 of them, 19 of which being confirmed or
candidate clusters.


\subsection{The IR photometric system and the adopted JHK magnitudes}

After a set of extensive tests of the different kind of magnitudes provided in
the PSC and XSC catalogues (e.g. isophotal, Point Spread Function fitting,
aperture), we decided to rely  on the aperture photometry in both cases.
Unfortunately, magnitudes  measured with the same aperture were not available
for both catalogues; hence our final choice was to use the aperture photometry 
within $r=4\arcsec$ for the PSC and $r=5\arcsec$ for the XSC.  All the 2MASS
magnitudes have then been transformed to the CIT photometric  system
\citep{CTI,eliasa,eliasb} using the color transformations provided  by
\citet{Carpenter}.

Fig.~2 shows the comparison between the 2MASS photometry and the previous NIR
photometries as adopted in the MC for the sources identified in the XSC (left
panels)  and in the PSC (right panels). In all cases the 
average absolute difference is $<0.1$ mag and the rms scatter around the mean
is $<0.2$ mag. No residual color effect has been detected. 

There are small systematic shifts determined from Fig.~2, probably due
to the difference in the adopted radius of the aperture. These have
thus been applied to the 2MASS photometry to put all of the measures
to the same photometric calibration adopted in the MC.

In Fig.~3 the JHK photometry of the total sample (obtained by  merging the
corrected magnitudes from the PSC and XSC) are compared  with the previous NIR
photometry from the MC. As can be seen, (a) the corrected magnitudes from both
the PSC and the XSC form a homogeneous set, and (b) for the vast majority of
the candidates the difference between corrected 2MASS photometry and previous
photometries is less then $\pm 0.2$ mag in all passbands, i.e., of the order of
the intrinsical uncertainties of pre-2MASS NIR measures. 
 The few objects displaying a difference larger that $\simeq 0.5$ mag
 has been checked by visual inspection of DSS images. All of them lie
 within $\sim 8-11 \arcsec$ of another source of similar or larger luminosity,
 hence their magnitude estimates may be influenced by the contamination of the
 close companions. 
The above procedure has provided $JHK$ photometry for 529 objects  for which no
previous NIR photometry was available. In particular, the final catalogue
contains homogeneous NIR data for 291 confirmed GCs, 412 candidates,  8
controversial objects, 58 confirmed galaxies, 23 confirmed stars and
3 H~{\sc ii}
regions. In conclusion, the NIR photometric dataset of the M~31 GC  candidates
has been nearly tripled by the present work.

\section{NIR-optical color-color diagrams:
A powerful tool for a preliminar classification}

   \begin{figure*}
   \centering
   \includegraphics[width=11.3cm]{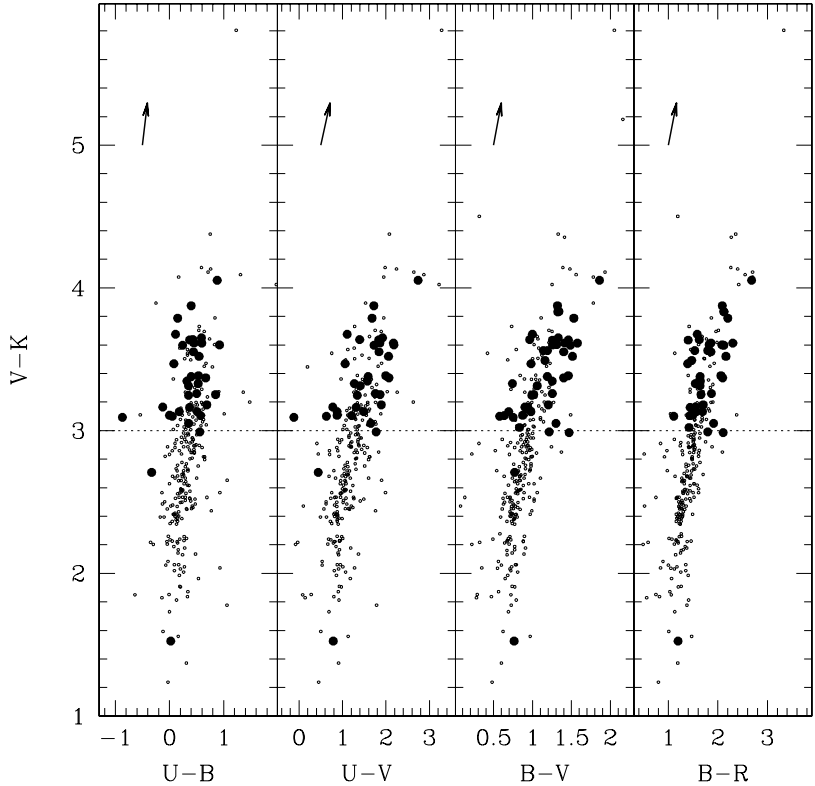}
      \caption{$V-K$ vs. various optical colors for the confirmed globular
      clusters (small points) and for the confirmed galaxies (filled circles). 
               }
          \label{FigVKGC}
  \end{figure*}

The knowledge of integrated colors offers a wealth of information on
any stellar population, and in particular, may be quite
useful for classification purposes. Here we show that
the available data on integrated colors may be used in many cases to
discriminate between star clusters and background galaxies thanks to
the large baseline provided by optical-infrared colors.

Various color-color diagrams with $V-K$ vs. optical colors are  presented in
Fig.~4 where {\it small points} are confirmed M~31 globulars,  and {\it filled
circles} are the confirmed galaxies that are present in our  Master Catalogue.
These galaxies should be typical of the background galaxies
that can contaminate the set  of candidate M~31 GCs. In each panel of
Fig.~4 (as well as in Fig.~5 below)  the arrow indicates the reddening vector
corresponding to $E(B-V)=0.1$ to illustrate quantitatively the effects of the
interstellar exctinction.

There are two remarkable features in Fig.~4:  (1) the confirmed GCs (points)
span a range of more than 3 magnitudes  in $V-K$, much larger than in optical
colors (also due to the larger impact of  the reddening in $V-K$);  (2) the
vast majority of confirmed galaxies (filled circles) has  $V-K\ge 3.0$, while
$\sim 50$\% of confirmed globulars are bluer than this  limit, reaching colors
as blue as $V-K\sim 1.2$. 
   \begin{figure*}
  \centering
   \includegraphics[width=11.3cm]{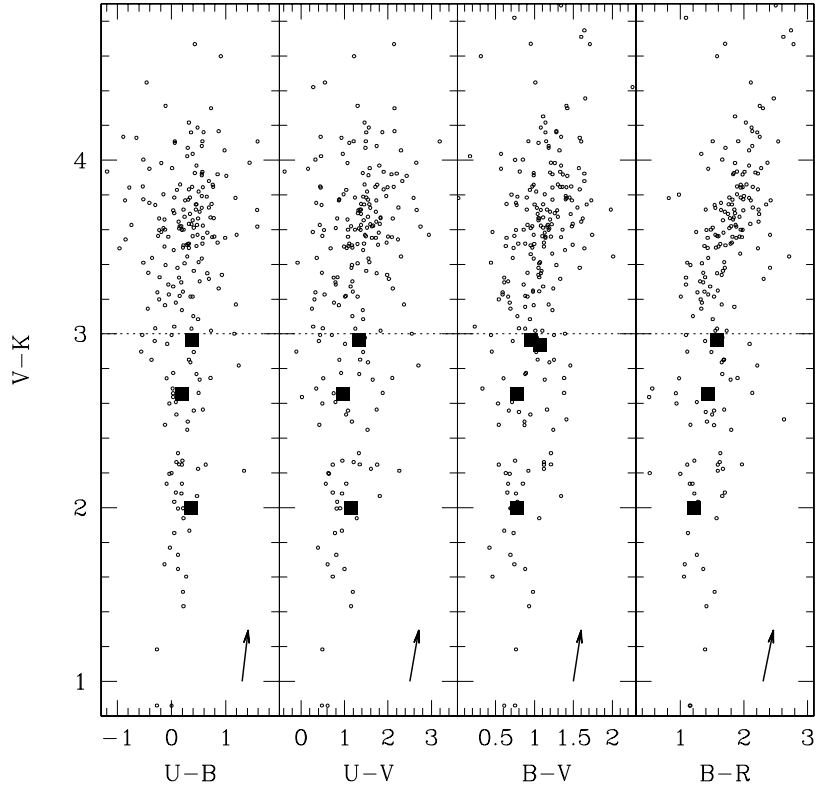}
     \caption{$V-K$ vs. optical colors for the not-yet-confirmed 
     candidates globular cluster (small open circles) and for controversial 
     objects that have been classified as clusters by some authors and as
     galaxies by others (filled squares).
               }
          \label{FigVKCand}
  \end{figure*}
%

Therefore the $V-K$ colors provide a useful discriminant to separate good
candidate clusters from background galaxies. This should lead to higher
efficiency in observational programs aimed at assessing the classification of 
clusters in M~31, since it is very unlikely that a candidate having $V-K< 3.0$
could be a misclassified background galaxy. 
On the other hand, foreground stars that may be erroneously classified as 
candidate M~31 globulars span the whole range of $V-K$ colors covered by 
confirmed clusters. Therefore, while the above criterion is very valuable for
the galaxy/cluster discrimination, it is ineffective in the star/cluster
separation.

It may also be useful to note that most of the clusters having  $V-K\ge 4.0$
lie in a region of the M~31 disk that seems to be affected by higher extinction
\citep{Barmbya}.  The isolated cluster at $V-K> 5.0$  is B037. 
Its very red color
is probably due to a color excess $E(B-V)>1.0$ mag \citep[in which case this
may be the brightest globular of the whole M~31 system, see][for discussion and
references]{Barmby02}, or to contamination by a bright and cold star 
accidentally superposed to the cluster. The confirmed galaxy at $V-K\sim 2.75$
is B136D  \cite[confirmed by][]{huchra91}, the one  at $V-K\sim 1.5$ is 
B390  \cite[classified as galaxies
by][]{racinea}.

Making use of the indications drawn from the plots presented in Fig.~4
for the
confirmed clusters and galaxies, one can apply the above criterion to get
useful hints on the nature of candidates whose  classification is uncertain. In
Fig.~5 we show the same color-color diagrams as in Fig.~4 for the
not-yet-confirmed candidate GCs (small open circles) in our Master Catalogue 
and for  the
four controversial objects (filled squares) classified as ``cluster'' by some
authors and as ``galaxy'' by others. It can be readily seen that most
candidates have colors consistent with those typical of background galaxies
($V-K\ge 3.0$), hence it can be reasonably  deduced that many of them are in
fact misclassified galaxies. On the other  hand, 83 candidates have $1.2\le
V-K<3.0$ and, according to the above  discussion, the contamination of this
sub-sample by background galaxies  should be negligible. A spectroscopic
campaign on these blue candidates  would probably be very rewarding in terms of
confirmation of {\em bona fide} clusters per observed candidate.
On the other
hand, it has to be recalled that the criterion introduces an obvious bias
against intrinsically red and/or highly reddened clusters, hence it cannot be
considered as a suitable method to select complete samples of candidate
clusters.

The controversial candidates G268, B409, G355 and B140 have been classified  as
galaxies by \citet{racinea}, based on high resolution ground based imaging,
while \citet{federicia,huchra91,Perrett} classified them as globular clusters
on the basis of spectroscopical observations.  All these candidates have
$V-K<3.0$, hence their $V-K$ colors seem to confirm the classification by
\citet{federicia} and \citet{Perrett}. For the other four controversial 
candidates not mentioned here, the controversy is between the classification as
a cluster or as a star, hence the doubt cannot be resolved with the presently
adopted criterion.

\section{The Revised Bologna Catalogue}

The MC described in Sect. 2, implemented with the new set of NIR
magnitudes obtained from 2MASS constitutes a new version of the
\citet{Battistinia} catalogue: the Revised Bologna Catalogue of
candidate GCs in M~31. The Revised Bologna Catalogue includes 337
confirmed GCs, 688 GC candidates, 10 objects with controversial
classification, 70 confirmed background galaxies, 55 confirmed stars
and 4 H~{\sc ii} regions. The whole catalogue is organized in
three different tables, all of them made available at the CDS in
electronic form.

Identifications and adopted coordinates are reported in Table 2, that
contains the following columns:

\begin{itemize}

\item $Column$ 1: Designation derived from the object
coordinates, following \citet{Barmbyc} (M31JHHMMSS+DDMMSS)

\item $Column$ 2: Designation: B and BxxxD from \citet{Battistini0,
Battistinia}, G from \citet{Sargent}, V from \citet{Vetesnik}, DAO from
\citet{Crampton}, NB from \citet{Battistinib}, AU from \citet{Auriere}, M from 
\citet{Mochejska}, BH from \citet{Barmbyc}, SH from \citet{Sharov83}, BA from
\citet{Baade})

\item $Column$ 3--12: Different identifications, respectively from 
\citet{Battistinia, Sargent, Vetesnik, Battistinib, Auriere, Crampton, Baade,
Sharov, Mochejska, Barmbyc}

\item $Column$ 13: Classification flag describing the nature of the entry  (1:
confirmed GC, 2: GC candidate, 3: controversial object, 4: confirmed  galaxy,
5: confirmed H~{\sc ii} region, 6: confirmed star)

\item $Column$  14: Classification from the original papers (\citet{Battistinia,
Mochejska, Barmbyc})

\item $Column$ 15:  Spectroscopic confirmation: C - cluster, G - galaxy,  S-
star, H - H~{\sc ii}  region

\item $Column$ 16:  Confirmation via high resolution imaging: C - cluster, G -
galaxy, S - star, H - H~{\sc ii}  region

\item $Column$ 17: comments and references

\item $Column$ 18 $and$ 19: RA, Dec (J2000) from the original catalogues or 
from DSS-2 or from HST/WFPC2 images (see Sect. 3). 

\item $Column$ 20: Sources of the adopted coordinates.

\end{itemize}

\bigskip\noindent The adopted photometric data have been collected and
are listed in  Table 3 for all the confirmed and candidate globular
clusters, and in Table 4 for all the objects confirmed to be galaxies,
stars or H~{\sc ii} regions.  The two tables have the following
columns:

\begin{itemize}

\item $Column$ 1: Designation, as in  Table 2

\item $Column$ 2:  Classification, as column 12 in Table 2

\item $Columns$ 3-10: $U,~B,~V,~R,~I,~J,~H,~K$ magnitudes (see Sect. 2)

\item $Column$ 14: Sources of the adopted optical data and NIR pre-2MASS.

\end{itemize}

The entries associated to the object listed in  Tab.~4 are reported at
the end of Tab.~2. In this way Tab.~2 can be easily combined with
Tab.~3 and Tab.~4 to obtain tables including positions and photometric
properties of ``candidate GC only'' and ``confirmed contaminant
objects only''.

\section{Summary and future prospects}

We have compiled a Master Catalogue of M~31 globular cluster
candidates including 1164 entries and we have searched the
counterparts of the MC objects in the NIR 2MASS databases. We have
found 776 2MASS counterparts, 693 of which correspond to confirmed or
candidate M~31 GCs. In this way we have provided integrated $J,~H,~K$
magnitude for 529 objects for which no previous NIR photometry was
available.

Moreover, we have collected and revised all the available
photometries, yielding a final adopted list of $UBVRIJHK$ magnitudes
as homogeneous as possible. The classification of a few candidates has
been assessed by inspection of HST images and a few controversial
cases have been reconsidered.  With all the newly collected material,
we have updated the original Bologna Catalogue \citep{Battistinia} so
obtaining the Revised Bologna Catalogue (RBC).  The RBC is the most
comprehensive list presently available of candidate M~31 clusters,
including 1035 confirmed and candidate GCs, 70 confirmed galaxies, 55
confirmed stars, and 4 H~{\sc ii} regions previously considered
possible GC candidates, for a total of 1164 entries.  The Bologna
Revised Catalogue is currently in use to prepare observational
programs aimed at increasing the sample of bona-fide M~31 globular
clusters and to study the integrated properties of the whole globular
clusters system of M~31.

It has also been shown that optical-infrared colors may be very
valuable in the discrimination between background galaxies and bona
fide globular clusters for $V-K<3.0$.  According to this criterion, we
can extract from the RBC a sub-sample of 83 objects whose
contamination by misclassified background galaxies is (probably) very
low.

\begin{acknowledgements}
We are very grateful to R.T. Rood for a critical reading of the original
manuscript and to an anonymous Referee for her/his useful comments and
suggestions.
This publication makes use of data products from the Two Micron All Sky 
Survey, which is a joint project of the University of Massachusetts and the
Infrared Processing and Analysis Center/California Institute of Technology,
funded by the National Aeronautics and Space Administration and the
National Science Foundation. Part of this work is based on observations made
with the NASA/ESA Hubble Space Telescope, obtained from the data archive at the 
Space Telescope Science Institute. STScI is operated by the Association of
Universities for Research in Astronomy, Inc., under NASA Contract NAS 5-26555.
This work  was supported by 
a fellowship (S.G.) from the {\it Consorzio Nazionale Astronomia 
ed Astrofisica--CNAA} and contributions from {\it 
MIUR-COFIN} and {\it Agenzia Spaziale Italiana--ASI}.
\end{acknowledgements}

\end{document}